\documentstyle[12pt]{article}

\newcommand{\nc}{\newcommand}
\def\frac#1#2{{\textstyle {#1 \over #2}}}

\nc{\beq}{\begin{equation}}
\nc{\eeq}{\end{equation}}
\nc{\beqa}{\begin{eqnarray}}
\nc{\eeqa}{\end{eqnarray}}
\nc{\lsim}{\begin{array}{c}\,\sim\vspace{-21pt}\\< \end{array}}
\nc{\gsim}{\begin{array}{c}\sim\vspace{-21pt}\\> \end{array}}

\input epsf
\newwrite\ffile\global\newcount\figno \global\figno=1

\def\writedef#1{}

\def\figin{\epsfcheck\figin}\def\figins{\epsfcheck\figins}
\def\epsfcheck{\ifx\epsfbox\UnDeFiNeD
\message{(NO epsf.tex, FIGURES WILL BE IGNORED)}
\gdef\figin##1{\vskip2in}\gdef\figins##1{\hskip.5in}% blank space instead
\else\message{(FIGURES WILL BE INCLUDED)}%
\gdef\figin##1{##1}\gdef\figins##1{##1}\fi}

\def\figinsert{}
\def\ifig#1#2#3{\xdef#1{fig.~\the\figno}
\writedef{#1\leftbracket fig.\noexpand~\the\figno}%
\figinsert\figin{\centerline{#3}}\medskip\centerline{\vbox{\baselineskip12pt
\advance\hsize by -1truein\center\footnotesize{  Fig.~\the\figno.} #2}}
\bigskip\endinsert\global\advance\figno by1}
\def\endinsert{}

\def\gev{\rm\,GeV}  
\def\mev{\rm\,MeV}

\def\&{and}

\def\nc#1#2#3{           {\it Nuovo Cim.  }{\bf #1}, #2 (19#3)}
\def\np#1#2#3{           {\it Nucl. Phys. }{\bf #1}, #2 (19#3)}
\def\pl#1#2#3{           {\it Phys. Lett. }{\bf #1}, #2 (19#3)}
\def\pr#1#2#3{           {\it Phys. Rev. }{\bf #1}, #2 (19#3)}
\def\prep#1#2#3{         {\it Phys. Rep. }{\bf #1}, #2 (19#3)}
\def\prl#1#2#3{          {\it Phys. Rev. Lett. }{\bf #1}, #2 (19#3)}

\begin{document}

\begin{titlepage}

\begin{center}
May, 1996      \hfill       YCTP-P9-96\\
%\hfill   preliminary draft\\
\vskip .5 in
{\Large \bf  Chiral Perturbation Theory, Large-$N_c$ \\
and the $\eta'$ Mass}
\vskip .3 in

%{    {\bf
%AUTHORS  }}

{{ Nick Evans\footnote{nick@zen.physics.yale.edu},
Stephen D.H. Hsu\footnote{hsu@hsunext.physics.yale.edu}
    and Myckola Schwetz\footnote{ms@genesis2.physics.yale.edu}
}}

   \vskip 0.3 cm
   {\it Sloane Physics Laboratory,
        Yale University,
        New Haven, CT 06511}\\
\end{center}

\vskip .5 in
\begin{abstract}
\noindent
Using chiral perturbation
theory and  the large-$N_c$ expansion,
we obtain expressions for the $\eta'$ mass
and $\eta - \eta'$ mixing in
terms of low-energy
chiral Lagrangian parameters.
This is accomplished through an intermediate
step of `matching' the topological susceptibility
in the large-$N_c$ and chiral Lagrangian
descriptions.
By inserting the
values of well-measured parameters we obtain
predictions involving the
the second order parameters
$L_6,L_7$ and $L_8$.
The prediction
for $L_6$ is quite restrictive even after allowing for
$1/ N_c$ corrections.

\end{abstract}
\end{titlepage}

%%%%%%%%%%%%%%%%%%%%%%%%%%%%%%%%%%%%%%%%%%%%%%%%%%%%%%%%%%%%%%%
\renewcommand{\thepage}{\arabic{page}}
\setcounter{page}{1}

\section{Introduction}
Chiral perturbation theory (CHPT) provides a useful description
of low-energy QCD in terms of the relevant light degrees of
freedom \cite{CHPT}. These light degrees of freedom are simply
the $N_f^2 - 1$
pseudo-Goldstone bosons (PGB's) which result from
the breaking of chiral symmetry:
$$SU(N_f)_L \times SU(N_f)_R \rightarrow SU(N_f)_V ~.$$
We do not include the $U(1)_A$ symmetry in the above
as it is violated by the axial anomaly \cite{anomaly}. The axial
anomaly has physical effects due to the presence in QCD of
field configurations with non-trivial winding number $\nu$.
Were this
not the case the chiral Lagrangian would have to be modified
to accomodate an additional (ninth) pseudo-Goldstone boson,
which has the quantum numbers of the $\eta'$.
It is well-known that this is what occurs in the large $N_c$ limit
-- the physical effects
of the anomaly are suppressed \cite{Witten}
\footnote{Despite this, the $\theta $-angle may still have physical
consequences in the large-$N$ limit \cite{Witten}. Also see \cite{Shur}
for a discussion of the possible large-$N_c$ behavior
of instantons. }.
In the large-$N_c$ limit the low-energy effective
Lagrangian contains the $\eta'$ as well as the  pions
\cite{NeffL}.

It might be regarded as a failure of the large-$N_c$
expansion that the $\eta'$ is so heavy -- $m_{\eta'} = 958 \mev$ --
relative to the other strange PGB's
such as the $K^{\pm}, K^0,
\bar{K}^0$ and
$\eta$. However, we will argue here that this is a consequence
of the smallness of the light quark masses and need not
imply a breakdown in large-$N_c$ thinking.
The point is that if one formally takes the limit
$N_c \rightarrow \infty$ with $m_{u,d,s}$ fixed,
it is easy to see that the $\eta'$ becomes roughly degenerate
with the strange PGB's (see section 2). However, it
is more appropriate for the real-world to consider a double
limit in which $N_c \rightarrow \infty$ and
$m_{u,d,s} \rightarrow 0$ simultaneously, in some
fixed ratio. Indeed, the breaking of the chiral $SU(3) \times SU(3)$
symmetry associated with the strange quark mass
 $\sim m_s / 4 \pi f_\pi \sim15 \%$
is still smaller than $1/ N_c$ ($N_c$=3) $\sim 30\%$.
As we will discuss below, after
taking this into
account it becomes clear that the $\eta'$ is heavy relative
to the PGB's because of the smallness of $m_s$ rather
than a failure of large-$N_c$. Using a simultaneous scaling
of $N_c$ and $m_q$ we can contemplate
a large-$N_c$ limit with the following hierarchy of masses:
\beq
\label{massh}
m_{\rho, baryon} ~>>~ m_{\eta'} ~>>~ m_{\pi,K,\eta}~.
\eeq
We note that a similar point of view was taken by
Leutwyler in \cite{HL}, where a bound on the light
quark mass ratios was derived using a double expansion
in large-$N_c$ and the quark masses.

We will take the viewpoint that large-$N_c$ gives
a qualitatively correct
picture of QCD dynamics, and exploit the tension
between the standard CHPT effective Lagrangian and the
one resulting from large-$N_c$. In particular, we will
utilize the special status of the $\eta'$ in determining
a quantity known as the topological susceptibility:
\beq
\label{sus}
{ { \langle~ \nu^2~ \rangle} \over {V} } ~=~ -~
{{1}\over{V Z}}~ \left(
{{ \partial^2 Z}\over{\partial \theta^2} } \right)_{\theta = 0}~,
\eeq
where $Z$ is the QCD partition function, $V$ the volume of
spacetime and $\nu$ is the winding number given by
\beq
\nu ~=~ \frac{1}{32 \pi^2} ~ \int_V d^4x ~ G \tilde{G}~   .
\eeq
In the large volume limit, $Z$ can be expressed
in terms of the vacuum energy density $\epsilon_0$:
\beq
\label{vol}
Z ~=~ exp{ (- V \epsilon_0 )}~.
\eeq
The problem is thus reduced to that of computing
the $\theta$ dependence of $\epsilon_0$.
This can be done either in CHPT, which is equivalent
to an expansion in the light quark masses $m_q$, or in
the large-$N_c$ effective Lagrangian, which involves
a double expansion in $1/N_c$ and $m_q$.
Equating the two results for the susceptibility will
yield a relationship at leading order in $N_c$
between $m_{\eta'}$ and
low-energy parameters appearing in CHPT.
An alternative way of viewing our calculation
is in terms of two different effective Lagrangians valid at
the two energy scales $\mu_1 \sim m_{\eta'}$
and $\mu_2 \sim m_{\pi,K,\eta}$ . At the lower scale
the $\eta'$ has been `integrated out'.
Requiring that both Lagrangians (one with the
$\eta'$, the other without) yield the same susceptibility
implies a non-trivial relation between their parameters.

In the following sections we will compute the topological
susceptibility in  both CHPT and Large-$N_c$ CHPT
to obtain the desired relation.
In the final section we
will present our results and some final comments.

\section{Large-$N_c$ Effective Lagrangian}

As previously mentioned, at large-$N_c$ the
effective Lagrangian must be modified to include
an additional light meson, the $\eta'$, which plays
the role of an additional PGB associated with
$U(1)_A$ \cite{Witten,NeffL}. It is convenient to
describe all $N_f^2$ PGBs in terms of a
$U(N_f)$ matrix field $U(x)$. The $\eta'$ is related
to the phase of the determinant of $U(x)$:
\beq
det U(x) = e^{-i \phi(x)}~
\eeq
with
\beq
\phi(x) = -{\sqrt{2N_f} \over F}~\eta' ~.
\eeq
The effective Lagrangian at large $N_c$ is
\beq
\label{LN}
{\cal L}_N ~=~ \frac{F^2}{4} tr( \partial_{\mu} U^{\dagger}
\partial^{\mu} U ) + \Sigma Re tr(M U^{\dagger})
- \frac{\tau}{2} (\phi - \theta)^2~,
\eeq
where the pion decay constant $F \sim {\cal O}(N_c^{1/2})$,
the chiral symmetry breaking scale $\Sigma \sim  {\cal O}(N_c)$
(related to the quark condensate in the chiral limit)
and $\tau \sim {\cal O} (1)$ (related to the topological
susceptibility of pure gluodynamics \cite{Witten}).
Note that the $\theta$ angle has been absorbed from
the mass matrix $M$ into a shift of $\phi$.
${\cal L}_N$ represents the leading order terms in the
effective Lagrangian in the formal counting scheme where
one takes the quark masses $m_q$ in the mass matrix $M$ to be
$\sim {\cal O} (N_c^{-1})$ and derivatives
$\partial \sim {\cal O} (N_c^{-1/2}) \sim m_{\pi}$.

Taking the quark mass matrix to be (we now restrict ourselves
to $N_f = 3$ and the $SU(2)$ isospin limit  $m_u=m_d=m$)
\beq
\pmatrix{m & 0 & 0 \cr 0 & m & 0 \cr 0 & 0& m_s}
\eeq
we obtain the following mass relations:
\beqa
\label{mrel}
m_{\pi}^2 ~&=&~ {2 \Sigma m \over F^2} \nonumber \\
m_{\eta}^2 ~&=&~ {2 \Sigma (m + 2m_s) \over 3 F^2} \nonumber \\
m_{\eta'}^2 ~&=&~ {2 \Sigma (2m + m_s) \over 3F^2}
+ {6 \tau \over F^2} \nonumber \\
\label{mrel}
m^2_{\eta \eta'} ~&=&~ {2 \sqrt{2} \Sigma (m - m_s) \over 3 F^2 } ~.
\eeqa
The first two equations are identical to the results
of CHPT, and
the last equation describes $\eta-\eta'$ mixing.
Note that both the anomalous and CHPT contributions
to $m_{\eta'}^2$ are formally ${\cal O} (1/N_c)$ in $N_c$
counting. As mentioned previously it is
possible to take the quark masses
$m_q \sim {\cal O} (N_c^{-1}) \rightarrow 0$
as $N_c \rightarrow \infty$ such that the anomalous
term continues to dominate -- i.e., $\tau >> \Sigma m_q$.
This is the limit that we will consider.
In real QCD where the $N_c \rightarrow \infty$ limit
is only approximate the anomalous contribution
to $m_{\eta'}$ is presumably not suppressed
below the CHPT contribution which is small
because of the quark masses. We argue therefore that
the large $N_c$ form of  (\ref{LN}) is appropriate to QCD.

The parameter $\tau$ can be
extracted at leading order
by requiring that the eigenvalues
of the $\eta-\eta'$ mass matrix correspond
to the phenomenologically observed values
of $547$ and $958~\mev$ respectively.  If we allow
CHPT and $1/N_c$  uncertainty of $50\%$ in the mass
relations (\ref{mrel}) (from, e.g., uncertainties
in $F_{\eta'}$, and subleading interactions
not in (\ref{LN}) which shift the mass relations (\ref{mrel}) etc.),
we obtain the result
\begin{equation} \label{taulimits}
(137 \mev)^4 <  \tau <  (202 \mev)^4~.
\end{equation}
In our calculations, we have used $F = 93 \mev$, the
allowed range of light quark masses $m_u \simeq m_d
\simeq 7 \mev$ and measured pion and kaon masses
to conlude that $m_s \simeq 150 \mev$ and
$\Sigma = {\cal O}(200\mev)^3$.

Note that  the above results are consistent
with the claim that $\tau \geq \Sigma m_s$ mainly due
to the smallness of $m_s$.

The next step is to compute the topological
susceptibility using ${\cal L}_N$. This requires
minimizing the potential energy at small $\theta$ angle
and differentiating
twice with respect to $i \theta$.
The $SU(N_f)$ isospin limit ($m = m_s$) of this problem
was analyzed previously by
Leutwyler and Smilga \cite{LS}.
In this case the global minimum of the potential
occurs when $U$ is a multiple of the
unit matrix $U = e^{-i \phi / N_f}$.
The vacuum energy is of the form
\beq
\label{vac}
\epsilon_0 ~=~ -N_f \Sigma m + \frac{1}{2} \theta^2
{\tau \Sigma m \over N_f \tau + \Sigma m} + {\cal O}(\theta^4) ~,
\eeq
and the topological susceptibility at $\theta = 0$ is
\beq
\label{ts}
{ { \langle~ \nu^2~ \rangle} \over {V} } ~=~
{\tau \Sigma m \over N_f \tau + \Sigma m} ~.
\eeq
%Note the behavior of ${ { \langle~ \nu^2~ \rangle} \over {V} }$.
In the limit we are considering, the denominator is
dominated by $\tau$, so
\beq
\label{ts1}
{ { \langle~ \nu^2~ \rangle} \over {V} }  ~\simeq~
{\Sigma m \over N_f} ~( 1 ~-~ {\Sigma m \over N_f \tau} ) ~.
\eeq
The second term is ${\cal O} (m^2)$, and one may wonder
whether it is consistent to retain it given the neglect of terms
of higher order in $( \Sigma M U )$ in ${\cal L}_N$. The point is that
due to an additional power of $\Sigma$
the term in (\ref{ts1}) is larger by a factor of $N_c$ than the
subleading corrections we have neglected. When we match
topological susceptibilites with the true low-energy theory
described by the $SU(N_f)$ chiral Lagrangian (no $\eta'$), this term
will fix a combination of $L_{6,7,8}$ at leading order in $N_c$.

It is important to note that the $\eta'$ and the PGB's play
a unique role in
determining ${ { \langle~ \nu^2~ \rangle} \over {V} }$.
The other particles not described by $U(x)$,
such as the $\rho$ meson, or the baryons $p,n, ...$ are
expected to have zero expectation value in the vacuum,
and hence do not contribute to the susceptibility even
if their interactions have some $\theta$-dependence.
This observation is independent of large-$N_c$, which
only guarantees that (\ref{ts}) is accurate to leading order.

The extension of (\ref{ts}) to the case of a heavy strange quark
is straightforward, and we obtain
\beq
\label{tss}
{ { \langle~ \nu^2~ \rangle} \over {V} } ~=~
{ \tau \Sigma {\cal M} \over \tau + \Sigma {\cal M} }~,
\eeq
where ${\cal M} = (1/m_u + 1/m_d + 1/m_s)^{-1}
= { m m_s / (m + 2m_s)}$. Note that this
reduces to $m / N_f$ where $N_f  = 3,2$ for
$m_s \rightarrow m$ and $m_s \rightarrow \infty$ respectively.
$\Sigma {\cal M}$ is approximately $\Sigma m$, which is
determined by the pion mass relation in (\ref{mrel}).

The range in $\tau$ given in (\ref{taulimits}) translates to the
following range in the topological susceptibility
\beq
\label{nulimits}
3.7 \times10^7 < { { \langle~ \nu^2~ \rangle} \over {V} } < 4.05
\times 10^7
\eeq
The uncertainty in $\tau$ does not lead to a large variation in
$ { \langle~ \nu^2~ \rangle} / {V}$.
There are, however, additional  $1/N_c$
corrections to the relation (\ref{tss}) which we have not yet
examined in detail. As mentioned previously, the
uncertainties represented in (\ref{nulimits}) are
due to $1/N_c$ corrections to the mass relations (\ref{mrel}).
However, subleading interactions which do not appear in
(\ref{LN}) may also alter the calculation of the
susceptibility.

There are two types of subleading corrections,
both formally of order $1/N_c$, which
we must consider.

\vskip  .5 cm

\begin{itemize}
\item{ Higher orders in $(\Sigma M U)$ which
induce corrections suppressed by ${\cal O} (m_q)$}.
The most important of these are of the form of
the $L_{6,7,8}$ interactions in (\ref{lchi}) below.
In large-$N_c$ counting $L_{6,7}$ are suppressed
by Zweig's rule so it is $L_8$ which is most important
\footnote{In the real world the central values of $L_{6,7}$
are roughly $ \sim  (1/N_c) L_8 $, which is at least
consistent with
the $1/N_c$ expansion approximation.}.

\item{${\cal O} (1/N_c)$ interactions of the $\eta'$ in ${\cal L}_N$
which violate the $U(N_f) \times U(N_f)$ flavor symmetries
but which preserve $SU(N_f)$ symmetries. These corrections
represent the fact that the $\eta'$ is only a PGB in the
large-$N_c$ limit. The most important of
these is of the form
\beq
\label{correction}
( \phi - \theta )^2 \Sigma Re tr(M e^{i \theta / N_f} U^{\dagger})~,
\eeq
where the coefficient is ${\cal O}(1/N_c)$.}

\end{itemize}

%\vskip  .3 cm

We can estimate the size of the uncertainties introduced
in the large-$N_c$ susceptibility by corrections of the above type.
Naively, the $L_{6,7,8}$ interactions will lead to a correction of
order ${\cal O} (m_s)$ which would be of order $15\%$.
However, the issue is somewhat more complicated than this.
The structure of $L_8$ is such that its effect on the
calculation of the susceptibility is equivalent to a shift in the
quark masses in (\ref{tss}): $m_i \rightarrow m_i (1 + {\cal O} (m_i) )$.
Hence only $m_s$ receives a non-negligible correction.
However, the strange quark mass has only a very
small effect on the susceptibility because it is so much more
massive than the u,d quarks. The parameter
${\cal M} = m  ( 1 +  {\cal O} (m / m_s) )$ in (\ref{tss}) depends
only weakly on $m_s$.
The $L_{6,7}$ operators, although suppressed
relative to $L_8$ by a power of $1/N_c$, have a
larger effect on the susceptibility.
An explicit calculation shows that $L_{6,7}$ shift
the susceptibility by ${\cal O} (m_s~1/N_c )$, leading
to a larger uncertainty than is given by the range
(\ref{nulimits}). This is the dominant subleading
correction. We will associate an additional
error of $\sim 5\%$ to this correction, leading to the
following range of susceptibility
(representing roughly  $10 \%$ error) :
\beq
\label{nulimits1}
3.5 \times10^7 < { { \langle~ \nu^2~ \rangle} \over {V} } < 4.3
\times 10^7~~.
\eeq

The second type of correction leads to a only a small
uncertainty in the susceptibility.
Let $\phi_0$ be the value of $\phi$ which minimizes
the vacuum energy $\epsilon_0$ in (\ref{vac}) at non-zero $\theta$:
\beq
\phi_ 0 = - ~{\theta  \over ( 1 + \Sigma {\cal M} /\tau ) } ~.
\eeq
We can then estimate the shift in the susceptibility by
substituting $\phi_0$
into (\ref{correction}) and differentiating twice with respect
to $\theta$. This leads to a correction which is suppressed by
a factor of
${\cal O} ( N_c^{-1} ~\Sigma {\cal M} / \tau )$, making it less
than $5 \%$ of (\ref{tss}).

\section{Chiral Lagrangian}

We next compute the topological
susceptibility using the standard
 $SU(N_f)$ chiral Lagrangian appropriate below the
$\eta'$ mass.
The heavy $\eta'$ has been `integrated-out' and its effects
are exhibited only in the chiral coefficients.
The susceptibility only depends on the potential of the
chiral Lagrangian. Terms with derivatives are irrelevant.

At ${\cal O} (m^2)$ the potentail is \cite{CHPT}
\begin{eqnarray}
\label{lchi}
{\cal V}_\chi ~=~ -~ \Sigma Re tr(M_\theta U^{\dagger}) ~&-&~
L_6 \frac{16\Sigma^2}{F^4} (Re tr  M_\theta U^{\dagger})^2 ~-~
L_7 \frac{16\Sigma^2}{F^4} (Im tr  M_\theta U^{\dagger})^2  \nonumber \\
{}~&-&~L_8  \frac{8\Sigma^2}{F^4} Re tr(M_\theta U^{\dagger}M_\theta
U^{\dagger})~,
\end{eqnarray}
where the mass matrix $M_{\theta}$ now incorporates the $\theta$-angle,
$M_{\theta} = M \exp{i \theta/N_f}$.
Here the $U(x)$ matrix field is restricted to unit determinant and is
parameterized as
\beq
U(x) ~=~ exp( i 2 \pi^a (x) T^a / F )~.
\eeq

The topological susceptibility is again found by minimizing the
potential for $U$ and taking the second derivative of $Z$
with respect to $(i \theta)$ at the minimum. We find
\begin{eqnarray}
\label{nuchi}
{ { \langle~ \nu^2~ \rangle} \over V} & =  & {2\over 9}
\left[ ~{\Sigma (2m + m_s) \over 2}  + {16 \Sigma^2\over F^4}
(~(L_6+L_7) (2 m + m_s)^2 + L_8 (2 m^2 + m_s^2)~) ~~~-   \right.\\
&& \nonumber  \\
& & \hspace{-0.8cm} \left. {{ 4 (m_s - m)^2 (~\frac{\Sigma }{2} +
\frac{16\Sigma^2}{F^4}
(~ (L_6+L_7) (2 m + m_s) + L_8 ( m + m_s) ~)~)^2}\over
{ \Sigma (m + 2 m_s) + \frac{16\Sigma^2}{F^4}
( 2 L_6  (2 m + m_s)( m + 2 m_s) + 4 L_7 (m_s - m)^2
+ 2 L_8 (m^2 + 2 m_s^2) )}} ~\
\right]. \nonumber
\end{eqnarray}
By equating this result, which is a function of
the second order parameters of the chiral
Lagrangian (\ref{lchi}), with the range (\ref{nulimits})
determined from the large $N_c$ expansion we can
impose constraints on the values of the parameters
$L_6,~L_7$ and $L_8$ in the chiral Lagrangian (\ref{lchi}).

The expression (\ref{nuchi}) has some interesting properties.
The leading order behavior is
\beq
{ { \langle~ \nu^2~ \rangle} \over V} ~=~ \Sigma {\cal M}~+~ \cdots,
\eeq
which agrees with the leading order large-$N_c$ result
(\ref{tss})
for $\tau >> \Sigma {\cal M}$.
Expanding (\ref{nuchi}) in powers of the light quark
masses, we can see that its leading dependence on $L_6$
is $\sim L_6 (\Sigma m) (\Sigma m_s / F^4)$, whereas the
leading dependence in $L_{7,8}$ is
$\sim L_{7,8} (\Sigma m) (\Sigma m / F^4)$.
This implies that the susceptibility is much
more sensitive to changes in $L_6$ than $L_{7,8}$.
When we perform the matching of susceptibilities below,
we will find that $L_6$ is more restricted than $L_{7,8}$.

We can derive further constraints on the
$L_{6,7,8}$ by matching higher derivatives
of the vacuum energy with respect to $i \theta$
in the large-$N_c$ and chiral descriptions. This
corresponds to matching $\langle \nu^4 \rangle$,
$\langle \nu^6 \rangle$, etc. Unfortunately, we have
checked that in the chiral Lagrangian description
$\langle \nu^4 \rangle$ is still only weakly dependent
on $L_{7,8}$. The leading dependence on $L_6$ is again
larger by a factor of
${\cal O}(m_s/m)$ than the dependence on $L_{7,8}$.
Thus we do not believe that additional constraints on
$L_{7,8}$ will be obtained by further matching.

\section{Matching}

A fit to the parameters $L_6,  L_7$ and $L_8$
from low energy data yields the following
one-sigma range of values
\cite{CHPT}:
\beqa
L_6 ~&=&~ -.0002 \pm .0003   \nonumber \\
L_7 ~&=&~ -.0004 \pm .0002   \nonumber \\
L_8 ~&=&~ +.0009 \pm .0003~~.
\eeqa

Upon matching the results for the topological susceptibility
in the large $N_c$ (\ref{tss}) and standard chiral lagrangian (\ref{nuchi}) we
obtain a more restrictive prediction for the volume in $L_6,  L_7$ and $L_8$
space
consistent with the observed meson masses upto the errors described above
in (\ref{nulimits1}).
The combined constraints are  shown in Fig. 1.
The parameter $L_6$ is well determined by the fit.
$L_7$ and $L_8$ are not so well constrained.
In Fig 2 and Fig.3 we project our limits onto the
$L_6-L_8$ and $L_6-L_7$ planes. Figure 2 shows
the allowed regions of $L_6-L_8$ with $L_7$
unconstrained and Figure 3 shows
the allowed regions of $L_6-L_7$ with $L_8$
unconstrained.

%\newpage

$\left. \right.$  \hspace{-0.3in}\ifig\prtbdiag{}
{\epsfxsize10.5truecm\epsfbox{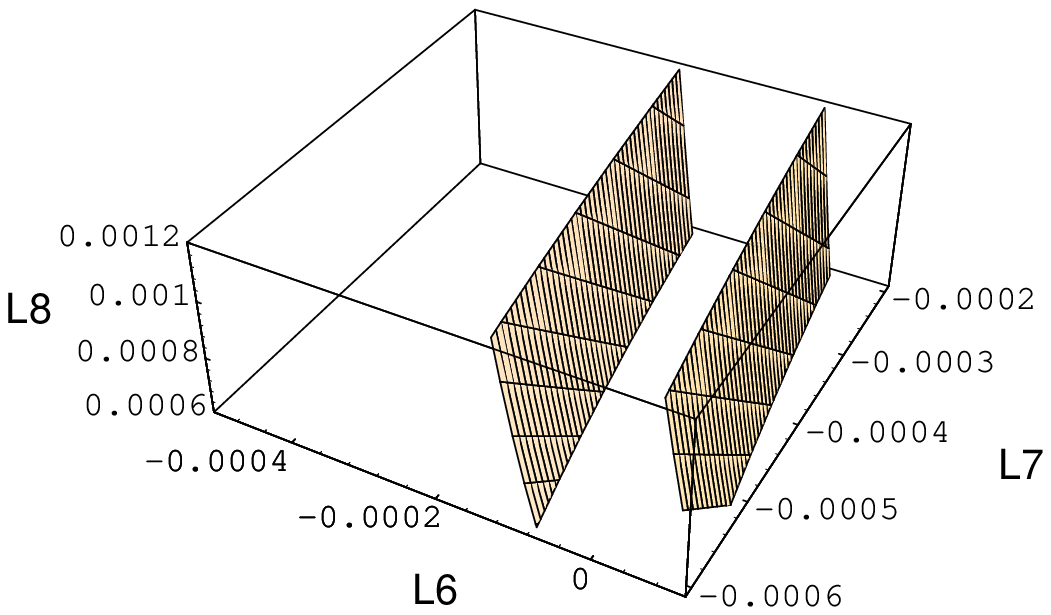}} \vspace{-1.85cm}
\begin{center} Fig.1: The region of $L_{6,7,8}$ space
consistent with the observed meson masses. \end{center}
\vspace{1cm}

$\left. \right.$  \hspace{-0.3in}\ifig\prtbdiag{}
{\epsfxsize8.5truecm\epsfbox{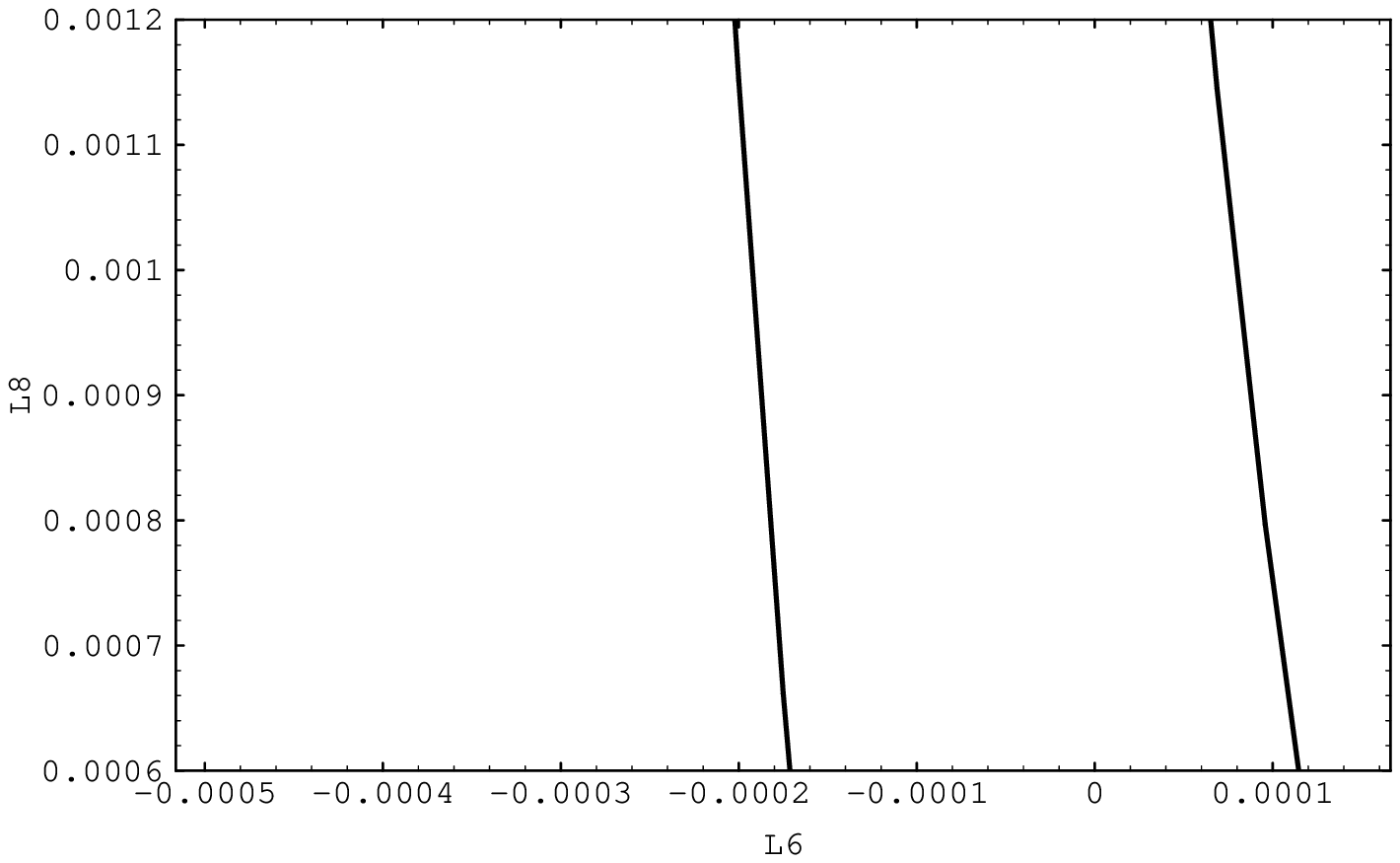}} \vspace{-1.85cm}
\begin{center} Fig.2: Projection of allowed region of parameter space
onto $L_6-L_8$ plane. \end{center}
\vspace{1cm}

$\left. \right.$  \hspace{-0.3in}\ifig\prtbdiag{}
{\epsfxsize8.5truecm\epsfbox{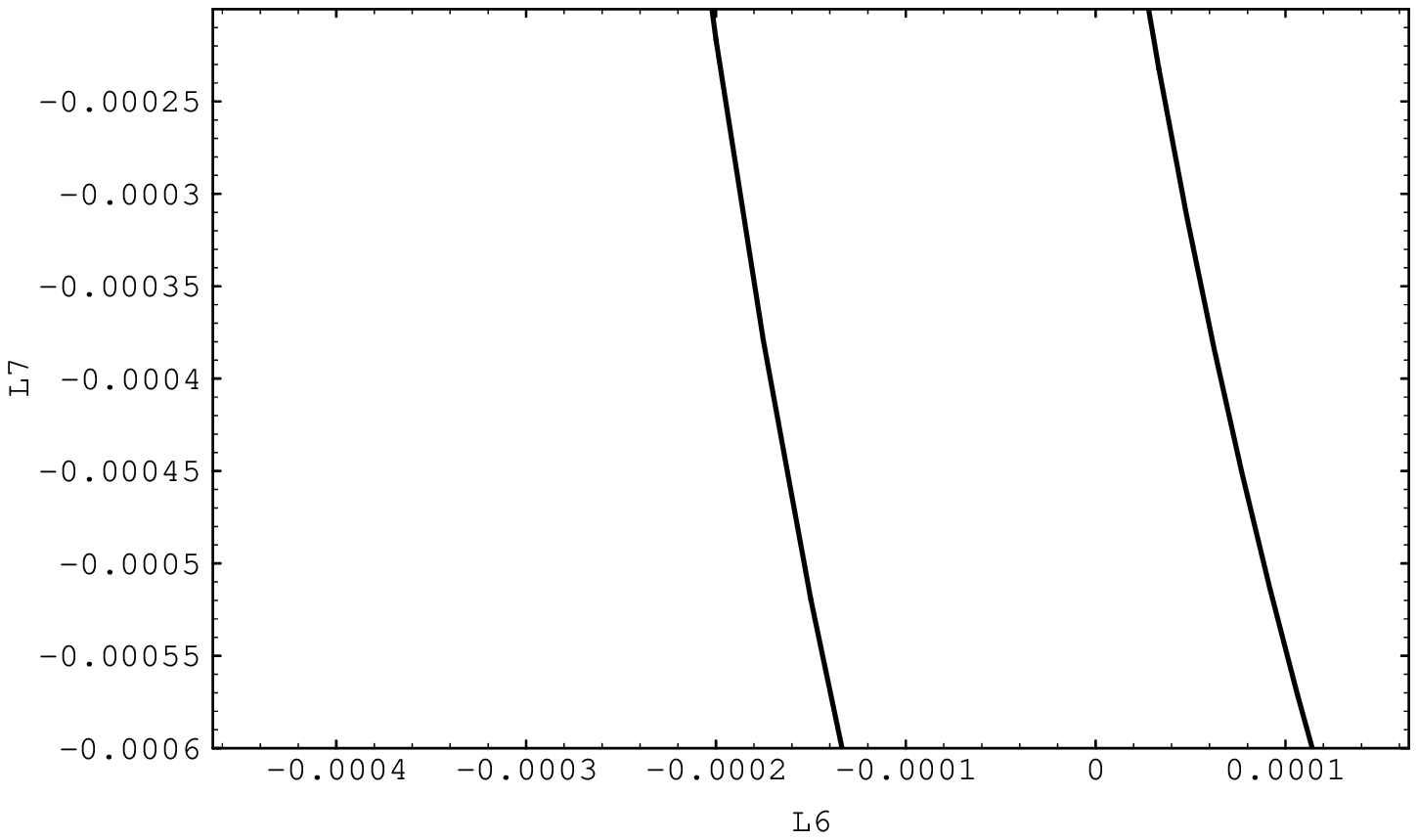}} \vspace{-1.85cm}
\nopagebreak
\begin{center} Fig.3: Projection of allowed region of parameter space
onto $L_6-L_7$ plane. \end{center}
\vspace{1cm}

\section{Discussion}

The $\eta'$ has certain special properties
in QCD as it is `almost' the
Goldstone boson associated with spontaneously
broken $U(1)_A$ symmetry. Within the large-$N_c$
expansion this `almost' can be made quantitative
and relations derived (see (\ref{tss})) between the
$\eta$ and $\eta'$ masses and the
topological susceptibility. The latter quantity measures the
sensitivity of the QCD vacuum energy to small
changes in $\theta$ at $\theta = 0$.
We can also compute the susceptibility directly
in chiral perturbation theory. The leading order
result in CHPT is precisely that obtained from
the large-$N_c$ Lagrangian if the $\eta'$ is taken
to be very massive ($\tau >> \Sigma {\cal M}$).
However, the subleading corrections in CHPT come
from $L_{6,7,8}$ whereas in the large-$N_c$ expansion
they depend on the ratio $\Sigma {\cal M} / \tau$.
We have exploited this observation to derive
what are essentially large-$N_c$ predictions for the
second order CHPT coefficients (particularly $L_6$)
in terms of the $\eta'$ mass which is taken as an input.
Our prediction for $L_6$ is restrictive even when
$1/N_c$ corrections are included. The central
value we predict is within the $1 \sigma$ region
previously extracted from low-energy data (also using
large-$N_c$) \cite{CHPT}.

A simple way of understanding our calculation is in terms
of effective Lagrangians valid at two scales,
$\mu_1 \sim m_{\eta'}$
and $\mu_2 \sim m_{\pi,K,\eta}$ . At the lower scale
the $\eta'$ has been `integrated out'.
The effect of virtual $\eta'$s appears in the
coefficients $L_{6,7,8}$ and is determined at leading order
in large-$N_c$. In this sense our calculation
is similar to that of \cite{resonance}, which estimated
the parameters $L_i$ from resonance exchange.

\vskip 1in
\newpage
\centerline{\bf Acknowledgements}
\vskip 0.1in
This work was supported under
DOE contract DE-AC02-ERU3075.

%\newpage

\end{document}